\documentclass[a4paper,10pt]{article}
\usepackage{a4wide,makeidx}                             
\usepackage[english]{babel}                             
\usepackage{amsmath,amsfonts,verbatim,graphicx,float} 

\begin{document}

\begin{center}
{\Large \bf Trigonometrically
extended Cornell potential and deconfinement }
\end{center}

\vspace{0.5cm}

\begin{center}
M.\ Kirchbach and C.\ B.\ Compean,\\
Instituto de Fis{\'{\i}}ca,
Universidad Autonoma de San Luis Potos{\'{\i}},\\
Av. Manuel Nava 6, Zona Universitaria,\\
S.L.P. 78290, M\'exico
\end{center}

\vspace{0.5cm}

\begin{flushleft}
{\bf Abstract:}
Non-perturbative methods of effective field 
theory such like Lattice QCD have allowed to establish   
connection between the QCD Lagrangian and quark potential models,
a prominent outcome being the Cornell (linear plus Coulomb)
potential. In being quite successful
in explaining properties of heavy flavor hadrons, be them 
quarkonia or baryons, this potential has been found less
spectacular in the description of the non-strange baryons, 
the nucleon and the $\Delta$. This  behavior indicates 
that the one-gluon exchange (giving rise to the inverse distance part) 
and the flux-tube interaction (responsible for the linear term)  
do not fully account for the complexity of the dynamics of
three light quarks. Very recently the Cornell potential which
is of infinite range
has been upgraded by us toward an exactly solvable trigonometric 
potential of finite range  that
interpolates between the inverse-distance and the infinite wells
while passing through a region of linear growth, a reason for which it
contains the inverse distance and linear potentials as first terms
in its Taylor series decomposition. 
We here first made the point that the upgraded Cornell potential  
can be viewed as the exact counterpart on a curved space to
a flat space $1/r$ potential, a circumstance that equips it by
two simultaneous symmetries, $SO(4)$, and $SO(2,1)$.
These allow to place the trigonometric confinement potential 
in the context of $AdS_5/CFT$ correspondence and thus to establish the link
of the algebraic aspects of the latter to QCD potentiology.
Next we report that the above potential when employed in 
the quark-diquark system,
provides a remarkably  adequate description of both nucleon and $\Delta $
spectra and the proton mean square charge radius as well, and
moreover implies an intriguing venue toward quark deconfinement 
as a shut-down of 
the curvature considered as temperature dependent.

\end{flushleft}



\vspace{0.5cm}

\section{ Introduction}

Significant progress in 
understanding  hadron properties has been reached    
through the elaboration of the connection between
the QCD Lagrangian and the potential models 
as deduced within the framework of effective field theories
\cite{Bali},\cite{Brambilla}
and especially through the  non-perturbative methods such as 
lattice simulations , the most prominent 
outcome being the linear plus Coulomb confinement potential \cite{Lattice},
\cite{Cornell}. The potentials derived from the QCD Lagrangian
have been most successful in the description
of heavy quarkonia and heavy baryon properties \cite{Nora_report}. 
Although the Cornell potential has found applications also in 
nucleon and $\Delta $ quark models \cite{varca}, 
the provided level of quality in the description of the non-strange sector 
stays  below the one reached for the heavy flavor sector.
This behavior indicates that  the one gluon exchange
(giving rise to  the Coulomb-like term) and of the flux-tube interaction
(associated with the linear part) do not fully  account for the complexity
of the dynamics of three light quarks.
Various improvements have been under consideration 
in the literature such as screening effects in combination with spin-spin
forces (see \cite{garz} and reference's therein).
 
Very recently, the Cornell potential has been upgraded by us in \cite{Quiry_2}
through its extension toward an exactly solvable 
(in the sense of the Schr\"odinger equation, or, 
the Klein-Gordon equation with equal scalar and vector potentials)
trigonometric quark confinement potential.
The latter has the property of
interpolating  between a Coulomb-like 
and an infinite well while passing through a region of linear growth,
and was shown to provide a remarkably good description of
the spectra of the non-strange baryons, the nucleon and the $\Delta$,
considered as quark-diquark systems, and of the proton charge radius as well
\cite{Quiry_2}. 

The potential under discussion is of infinite depth and has only bound states
which diagonalize the  $SO(4)$ Casimir operator of the 
four-dimensional angular momentum. 
Alternatively, same states diagonalize also, under certain conditions 
to be discussed below, the three 
dimensional pseudo-angular momentum, the Casimir operator of $SO(2,1)$. 
The first symmetry is embedded in the 
de Sitter group, $SO(1,4)$, which by-itself is a subgroup of 
the conformal group
$SO(2,4)$ according to the reduction chain,
\begin{eqnarray}
SO(2,4)\supset SO(1,4)&\supset& SO(4)\supset SO(3)\supset SO(2),
\nonumber\\
&& K\qquad \qquad l\qquad\quad  m.
\label{conf}
\end{eqnarray}
Here, $K$, $l$, and $m$ are the respective four--, three- and two-dimensional
angular momenta, $l=0,1,2,..K$, and $m=-l,..0, ...+l$.
Alternatively, the $SO(2,1)$ symmetry is part of the reduction chain
of the anti-de Sitter group, $SO(2,3)$, 
\begin{eqnarray}
SO(2,3)&\supset& SO(2,1)\supset SO(2),\nonumber\\
        && j \qquad \qquad\quad  m^\prime .
\label{Ads}
\end{eqnarray}
Here, $j$ is the so called Bargmann index that fixes the eigenvalues
of the pseudo-angular momentum, ${\mathcal J}^2=J_x^2+J_y^2 -J_z^2$
as ${\mathcal J}^2|jm^\prime \rangle =j(1-j)|jm^\prime \rangle $, and 
$m^\prime $ is 
the  $J_z$ eigenvalue, $J_z|jm^\prime \rangle =m^\prime |jm^\prime\rangle$, 
unlimited from above  according to $m^\prime =j, j+1, j+2, ...$.
The group symmetries under discussion appear within the context of
a Schr\"odinger equation written in different variables 
\cite{Wybourne},\cite{Marcelo}. 
\begin{quote}
The first goal of the present contribution is to draw attention to the fact
that the $SO(4)$ symmetry of the potential under discussion is well understood
in terms of the solution to the Laplace-Beltrami equation on
a three-dimensional sphere of constant radius, $S^3_R$, 
embedded in a four-dimensional flat Euclidean space, $E_4$, or. alternatively,
in an Euclidean anti-de Sitter space.
\end{quote}
The two symmetries are especially interesting in connection with a development
initiated by  \cite{AdS} according to which a
maximal supersymmetric Yang-Mills conformal field  theory (CFT)
in four dimensional Minkowski space is equivalent to a 
type IIB closed superstring theory 
in ten dimensions described by the product manifold
$AdS_5\times S^5$. More recently,  the correspondence between 
string theory in ten dimensional anti-de Sitter 
space and $SO(4,2)$ invariant conformal Yang-Mills theories, so called
$AdS_5/CFT$ correspondence, 
has been adapted in \cite{Brodsky} to the description
of hadron properties.

\begin{quote}
We here  emphasize  
that the $SO(4)$ and $SO(2,1)$ symmetries of the ``curved'' potential 
are in line with the algebraic aspects of the $AdS_5/CFT$ 
correspondence as visible from eqs.~(\ref{Ads}) and (\ref{conf})
thus establishing a link of the latter to QCD potentiology.
\end{quote}
However, it is important to be aware of the fact that the algebraic
$AdS_5/CFT$ criteria alone
are not sufficient to fix uniquely the potential.
One  has to complement them by the requirement on compatibility with 
the QCD Lagrangian too, a condition which
imposes severe restrictions on the allowed potential shapes.
\begin{quote}
Our next point is that it is precisely
the trigonometrically extended Cornell potential, 
employed in baryons as quark-diquark potential, 
the one that meets best both the $AdS_5/CFT$ and QCD criteria
and provides the link between them.
Predicted degeneracy patterns and level splitting are such
that none of the observed  $N$ states 
drops out from the corresponding systematics
which also applies equally well to the $\Delta $ 
spectra (except accommodation of the hybrid $\Delta (1600)$). 
The scenario provides
a remarkable description of the  proton charge electric
form-factor too and moreover implies a deconfinement mechanism
as a shut-down of the curvature considered as temperature dependent.
\end{quote}

\section{ The ``curved'' Cornell potential}
Around 1941 Schr\"odinger had the idea to solve  
the quantum mechanical Coulomb problem in the cosmological context of 
Einstein's closed universe, i.e. on the three dimensional  (3D) hypersphere, 
$S_R^3$,  of a constant radius $R$ \cite{Schr40}.
In order to construct the counterpart of the flat-space inverse 
distance potential on $S_R^3$ Schr\"odinger had to solve the 
corresponding four-dimensional Laplace-Beltrami equation. 
In result, after some straightforward
algebra, he found the following potential

\begin{eqnarray}
{\mathcal V}(\chi )&=&
c\cot \chi +\kappa \frac{\hbar^2}{2\mu }\frac{l(l+1)}{\sin^2\chi },
\quad \kappa =\frac{1}{R^2}.
\label{chi_pot}
\end{eqnarray} 
Here $\chi $ stands for the second polar angle in $E_4$,
and the $\csc^2$ term describes the centrifugal barrier
on $ S_R^3$.
Schr\"odinger's prime result, the presence of curvature provokes 
that the orbiting particle appears confined 
within a  trigonometric potential of infinite depth
and the hydrogen spectrum shows only bound states.
This is a very interesting situation in so far as in $E_4$
the r\'ole of  the radial coordinate of infinite range in 
ordinary flat space, $0<{ r}< \infty $, has been taken by the
angular variable, $\chi$, of finite range. In other words, 
while the harmonic potential in $E_3$ is a  central one, in
$E_4$ it is non-central. Moreover, 
\begin{quote}
the inverse distance potential of finite
depth in $E_3$ is converted to an infinite barrier and therefore to a
confinement potential in the higher dimensional $E_4$ space, 
a property  of fundamental importance throughout the paper. 
Stated differently, confinement phenomena can be associated with
infinite barriers due to curvature. 
\end{quote}

 An especially interesting observation
was that the $O(4)$ degeneracy of the levels observed in the 
flat space $H$ atom spectrum was preserved by
the curved space spectrum too in the sense that
also there the levels could be labeled by the standard atomic indices
$n$, $l$, and $m$, and the energy depended on $n$ alone.

The simplest barrier of such a nature is
the manifestly $O(4)$ symmetric $S_R^3$ centrifugal barrier,
$U_l(\chi ,\kappa) =\kappa \frac{\hbar^2 }{2\mu }\frac{l(l+1)}{\sin^2 \chi }$. 
Indeed,
the angular part,  $ \Box_{\Omega}   $, 
of the four-dimensional  
Laplacian, $\Box$,  relates to the operator of the
four-dimensional angular momentum, 
here denoted by ${\mathcal K}^2$,
\footnote{The analogue on the two-dimensional sphere of a constant radius
$r=a$  is the well known relation
${\vec \nabla}^2 =-\frac{1}{a^2}L^2 $.} according to,
$\Box_\Omega =-\frac{1}{R^2} {\mathcal K}^2 $,
whose action on the states is given by \cite{Kim_Noz}
\begin{equation}
{\mathcal K}^2 \vert K l m \rangle = K(K+2)
\vert K l m \rangle.
\label{Casimir_O4}
\end{equation}
 Therefore, the corresponding energy spectrum has to be,
$
E_K^{(c=0)}(\kappa )=\kappa \frac{\hbar ^2}{2\mu } 
K(K+2)$. When cast in terms of  $n=K+1$,
the latter spectrum becomes
$
E_n^{(c=0)}(\kappa )=\kappa \frac{\hbar ^2}{2\mu } 
\left( n^2-1\right)$,
which coincides in form (though not in degeneracies) with the spectrum of  a 
particle confined within
an infinitely deep spherical quantum-box well.
Then $n$ acquires meaning of  principal quantum number.
The complete solutions of eq.~(\ref{Casimir_O4}) especially 
on the unit hypersphere are text-book knowledge  
\cite{Kim_Noz} ,\cite{Tjon} and 
are the well known hyper-spherical harmonics,
$
|Klm>=Z_{Klm}(\chi, \theta, \varphi )= {\mathcal S}_{Kl}(\chi ,\kappa=1 )
Y_l^m(\theta ,\varphi)$,
where ${\mathcal S}_{Kl}$ contain the Gegenbauer polynomials.\\

In general, 
various potentials in conventional flat $E_3$ space
can be constructed as images to  ${\mathcal V}( \chi) $ in 
eq.~(\ref{chi_pot}).
Their explicit forms are determined by  the choice of 
coordinates on $S_R^3$ which shape the line element, $\mbox{d}s$. 
The general expression of the line element in the space under 
consideration and in hyper-spherical coordinates, 
$\Omega =\lbrace\chi, \theta, \varphi \rbrace $, reads
\begin{eqnarray}
\mbox{d}s^2&=&\frac{1}{\kappa}\lbrack \mbox{d}\chi^2 +
\sin^2\chi (\mbox{d}\theta ^2
+\sin^2\theta \mbox{d}\varphi^2)\rbrack.
 \label{metrics_chi}
\end{eqnarray}
Upon the variable substitution, $\chi = f\left( r\right)$, restricted to
$0 \leq f(r)\leq \pi$, 
eq.~(\ref{metrics_chi})
takes the form 
\begin{eqnarray}
\mbox{d}s^2&=&
D^2(r,\kappa )\frac{\mbox{d}r^2}{r^2} + {\mathcal R}^2 (r,\kappa )
(\mbox{d}\theta^2 +
\sin^2\theta \mbox{d}\varphi ^2),
\label{metrics_r}
\end{eqnarray}
where $D(r,\kappa )\equiv \frac{ r}{\sqrt{\kappa}} f^\prime (r)$, 
and  ${\mathcal R}(r, \kappa )\equiv \frac{\sin f(r)}{\sqrt{\kappa}}$ 
are usually referred to as 
``gauge metric tensor'' and ``scale factor'', respectively \cite{Ismst}.
Changing variable correspondingly in the associated Schr\"odinger equation
is standard and  various choices for  $f(r)$  give rise to a variety 
of radial equations in ordinary
flat space with effective potentials which are not even necessarily central.
All these equations, no matter how different that may look,
are of course equivalent, they have same spectra,
and the transition probabilities between the levels  
are independent on the choice for $f(r)$.
Nonetheless, some of the scenarios provided by the different choices for $f(r)$
can be more efficient in the description
of particular phenomena than others.

\begin{quote}
Precisely here lies the power of the curvature concept
as the common prototype of confinement phenomena of different disguises.
In the following we shall present one typical example for $f(r)$.
\end{quote}

\underline{\it 2.1 The $D(r,\kappa )=\pi r  $ 
gauge and the upgraded Cornell potential.} An especially simple and 
convenient parametrization of the $\chi$ variable  in terms of  $r$, 
also used by Schr\"odinger \cite{Schr40} and corresponding to the $D(r)=\pi r$
gauge  is
\begin{equation}
\chi =\frac{r}{R}\pi \equiv \frac{r}{d}, \quad d=\frac{R}{\pi }, \quad 
\frac{r}{R}\in [0, 1 ], \quad \kappa \to \widetilde{\kappa}= \frac{1}{d^2}.
\label{Schr_choice}
\end{equation} 
Here, the length parameter $d$ assumes the r\'ole of 
rescaled hyper-radius.  Correspondingly,
in this particular gauge, the place of the
genuine curvature, $\kappa =1/R^2$, is taken  by  
the rescaled one, $\widetilde{\kappa}=1/d^2$.
Setting now  $c=-2G\sqrt{\widetilde{\kappa} }$,  
amounts to the following radial Schr\"odinger equation,
\begin{eqnarray}
{\Big[} 
-\widetilde{\kappa }\frac{\hbar ^2}{2\mu}\frac{\mbox{d}^2 }{\mbox{d}
\left( r\sqrt{\widetilde{\kappa}}\right)  ^2}
&+& {\mathcal V} \left(r\sqrt{\widetilde{\kappa }},
\widetilde{ \kappa} \right){\Big]}  
X\left(r\sqrt{\widetilde{\kappa}},\widetilde{\kappa} \right)
=\left(E(\widetilde{\kappa} ) +\frac{\hbar^2}{2\mu}\widetilde{\kappa} \right) 
X\left(r\sqrt{\widetilde{\kappa}}, \widetilde{\kappa}  \right),\nonumber\\
{\mathcal V} \left(r\sqrt{\widetilde{\kappa} },
\widetilde{\kappa} \right)  &=&
\widetilde{\kappa} \frac{\hbar^2}{2\mu } 
\frac{l(l+1)}{\sin^2 \left( r\sqrt{\widetilde{\kappa} }\right) }
-2G\sqrt{\widetilde{\kappa}}\cot \left( r\sqrt{\widetilde{\kappa} }\right).
\label{ros_morse}
\end{eqnarray}
A similarly shaped  potential is managed by SUSYQM \cite{Levai}
under the name of Rosen-Morse I. The essential difference 
between Rosen-Morse I and ${\mathcal V}(r\sqrt{\widetilde \kappa },
\widetilde \kappa )$  is the suppression of the curvature in the former 
and its treatment as a central potential in ordinary flat space.
Instead, we here shall treat ${\mathcal V}(r\sqrt{\widetilde \kappa },
\widetilde \kappa )$ consequently as a ``curved''
and maintain intact the $S_R^3$ volume in all integrals to appear.

It is now quite instructive to expand ${\mathcal V}(r\sqrt{\widetilde \kappa },
\widetilde \kappa )$ in a Taylor series.
In so doing, one finds the following approximation,
\begin{eqnarray}
-2G\sqrt{\widetilde{\kappa}}\cot r\sqrt{\widetilde{\kappa} }
+ \widetilde{\kappa} \frac{\hbar^2}{2\mu } 
\frac{l(l+1)}{\sin^2 \left( r\sqrt{\widetilde{\kappa} }\right) }
&\approx&
-\frac{2G}{r} +\frac{2G\widetilde{\kappa}}{3} r
+ \frac{\hbar^2}{2\mu } 
\frac{l(l+1)}{r^2 },
\label{crnl}
\end{eqnarray}
with $\widetilde{\kappa}=\frac{1}{d^2}=\frac{\pi^2}{R^2}$.
\begin{quote}
Therefore, the finite range potential,
 ${\mathcal V}(r\sqrt{\widetilde{\kappa}},\widetilde{ \kappa} )$,
is the  exactly solvable 
{\it \underline{t}rigonometric  \underline{e}xtension
to the infinite range {\underline{C}ornell}\/} potential, to be
referred to  as TEC potential.
\end{quote}

Besides Schr\"odinger, eq.~(\ref{ros_morse})
has  been solved by various authors using different schemes.
The solutions obtained in \cite{Stevenson} are built on top 
of Jacobi polynomials of imaginary arguments and parameters that are
complex conjugate to each other,
while ref.~\cite{Vinitski} expands the
wave functions of the interacting case in the free particle basis.
The most recent construction in our previous work  \cite{Quiry} instead
relies upon real Romanovski polynomials. In the $\chi $ variable
in eq.~(\ref{Schr_choice})
our solutions to eq.~(\ref{ros_morse}) take the form, 
\begin{eqnarray}
X_{(K l)}(\chi , \widetilde{\kappa }  ) =  N_{( K l) } 
\sin^{K+1} \chi  e^{-\frac{b\chi }{K+1}  }
R_{K-l }^{(\frac{2b}{K+1 },-( K+1) )}\left(\cot \chi  
\right),
&\quad&  b=\frac{2\mu G}
{\sqrt{\widetilde{\kappa}}\hbar^2}.
\nonumber\\
K=0,1,2,...,\quad  l=0,1,...,K ,&&
\label{Rom_pol}
\end{eqnarray}
where  $N_{(K l)}$ is a normalization constant. 
The 
$R_{n}^{(\alpha, \beta )}(\cot \chi )$ 
functions are the non-classical 
Romanovski polynomials \cite{routh,rom} which are defined by the
following Rodrigues formula,
\begin{eqnarray}
R_{n}^{(\alpha, \beta )}(x)&=&e^{\alpha  \cot^{-1} x} (1+x^2)^{-\beta +1}
\nonumber\\
&&\times \frac{\mbox{d}^n}{\mbox{d}x^n}e^{-\alpha  
\cot^{-1} x} (1+x^2)^{\beta -1 +n},
\label{Rodrigues}
\end{eqnarray}
where $x=\cot r\sqrt{\widetilde{\kappa}} $ 
(see ref.~\cite{raposo} for a recent review).

The energy spectrum of ${\mathcal V}\left(r\sqrt{\widetilde{\kappa} },
\widetilde{\kappa } \right)$ 
is calculated as
\begin{equation}
E_{K}(\widetilde{\kappa} )=-\frac{G^2}{\frac{\hbar^2}{2\mu}}   
\frac{1}{(K+1)^2}
+ \widetilde{\kappa} \frac{\hbar ^2}{2\mu }( (K+1)^2-1), \quad l=0,1,2,...,K.
\label{enrg_cot}
\end{equation}
Giving $(K+1)$ the interpretation of a principal quantum number
 $n=0,1,2,...$ (as in the $H$ atom),
one easily recognizes that the energy in eq.~(\ref{enrg_cot}) is
defined by the Balmer term and its inverse of opposite sign, thus
revealing $O(4)$ as dynamical symmetry of the problem.
Stated differently, particular  levels 
bound within different  potentials (distinct by the values of $l$) 
carry same energies and align to levels 
(multiplets) characterized, similarly to the free case in 
eq.~(\ref{Casimir_O4}), by the four dimensional angular momentum,
$K$. The  $K$-levels belong to the irreducible  
$O(4)$ representations of the type  $\left(\frac{K}{2},\frac{K}{2} \right)$. 
When the confined particle carries spin-1/2, as is
the case of electrons in quantum dots, or quarks in baryons, 
one has to couple the spin, i.e. the
$\left(\frac{1}{2},0 \right)\oplus \left(0,\frac{1}{2} \right)$ representation,
 to the previous multiplet,
ending up with the (reducible) $O(4)$ representation
\begin{equation}
\vert Klm, s =\frac{1}{2}\rangle = \left(\frac{K}{2},\frac{K}{2} \right)
\otimes \left[
   \left(\frac{1}{2},0 \right)\oplus \left(0,\frac{1}{2} \right)\right].
\label{cluster-K}
\end{equation}
This representation contains  $K$ parity dyads
and a state
of maximal spin, $J_{\mathrm{max}}=K+\frac{1}{2}$, without
parity companion and  
of either positive ($\pi =+$) or, negative ($\pi =-$) parity,
\begin{equation}
\frac{1}{2}^\pm , ...,\left( K-\frac{1}{2}\right)^\pm, 
\left( K+\frac{1}{2}\right)^\pi \in \vert Klm, s=\frac{1}{2}\rangle .
\label{filon}
\end{equation}
As we shall see below this scenario turns to be remarkably adequate for
the description of non-strange baryon structure.

\section{  $S_R^3$ potentiology:The baryons}

\underline{ 3.1 \it The nucleon spectrum in the $SO(4)$ symmetry scheme. }
The spectrum of the nucleon continues being enigmatic despite 
the long history of the respective studies (see refs.~\cite{Lee}, 
\cite{Afonin} for recent reviews).
Unprejudiced inspection of the data reported by the Particle
Data Group \cite{PART} reveals a systematic degeneracy  of the
excited states of the baryons of the best coverage, the
nucleon $(N)$ and the $\Delta (1232)$. 
Our case is that
\begin{itemize}
\item  levels and level splittings of the nucleon and $\Delta $ spectra
match remarkably well the spectrum in eq.~(\ref{enrg_cot}).
\end{itemize}

\begin{enumerate}
\item \underline{The $N$ and $\Delta$ spectra:}
The measured nucleon resonances with masses below $2.5$ GeV
fall into the three $K= 1,3,5$ levels in eq.~(\ref{filon})
with only the two 
$F_{17}$ and $H_{1,11}$ states still ``missing'', 
a systematics anticipated earlier by one of us (M.K.) 
in refs.~\cite{MK-97} on the basis of pure algebraic considerations.
Moreover, the level splittings follow with an amazing accuracy 
eq.~(\ref{enrg_cot}). The nucleon spectrum in the quark-diquark 
picture of internal structure, shown in Fig.~1, is fitted by the following 
parameters of the ``curved'' potential in  eq.~(\ref{ros_morse}), 
\begin{eqnarray}
&\mu= 1.06 \,\, \mathrm{fm}^{-1}\ ,\quad G=237.55\  
\mbox{MeV}\cdot\mbox{fm} \ ,\quad d=2.31 
\ \mathrm{fm}.
\label{parameters}
\end{eqnarray}
\begin{figure}[b]
\center
\includegraphics[width=10.7cm]{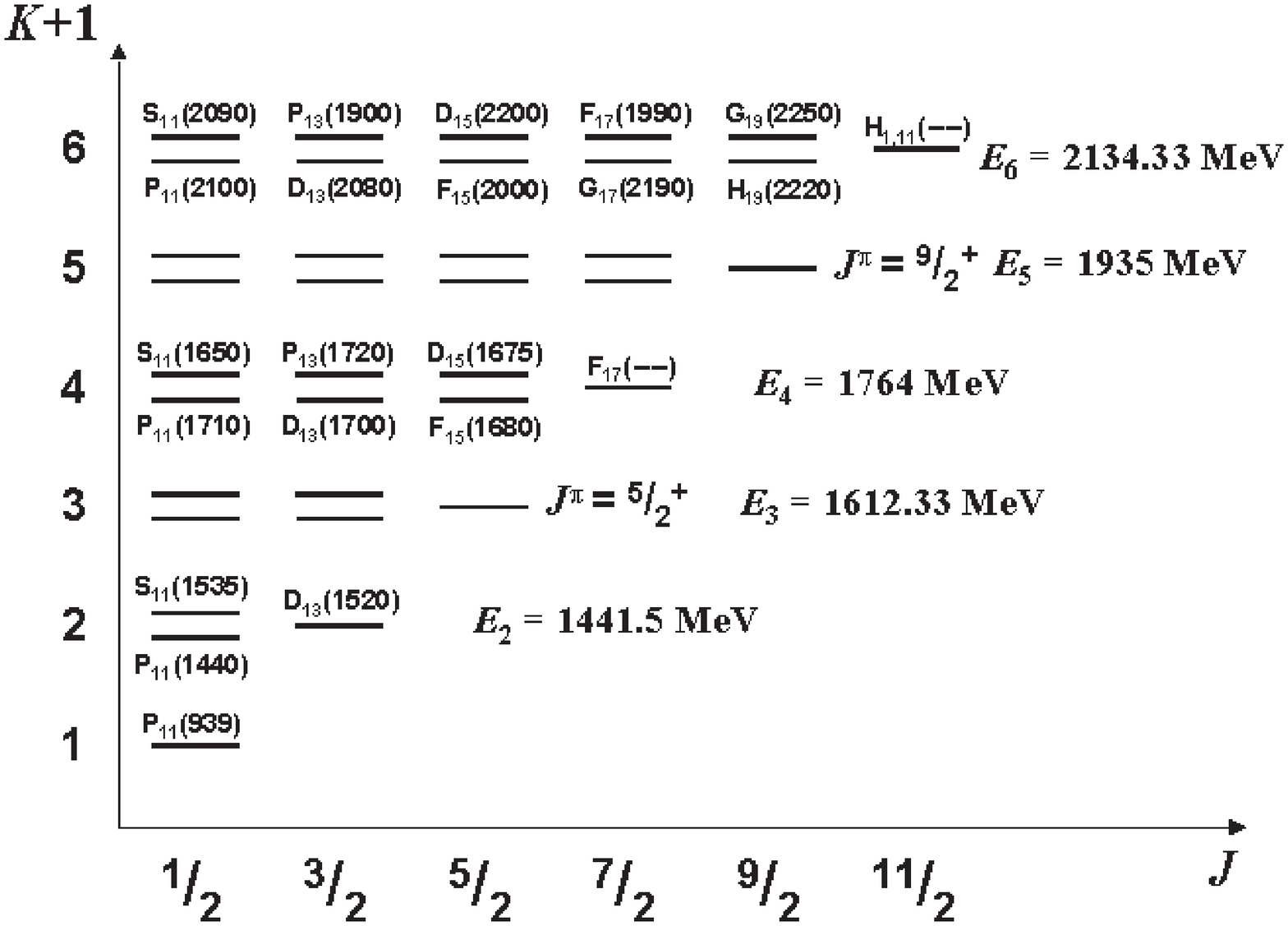}
\caption{
Assignments of the reported $N$ excitations to
the $K$ levels of the $S_R^3$ potential,
 ${\mathcal V}(r\sqrt{\widetilde{\kappa}},\widetilde{ \kappa} )$,
in eq.~(\ref{ros_morse}), taken as the quark-diquark confinement potential.
The potential parameters are those from eq.~(\ref{parameters}).
Double bars represent parity dyads, single bars the unpaired states of maximum
spin. The notion $L_{2I,2J}(--)$ has been used  for resonances  ``missing'' 
from a level. 
The model predicts two more levels of maximal spins $J^\pi =5/2^+$, 
and  $J^\pi =9/2^+$, respectively, which are completely  ``missing''. 
In order not to overload the figure with notations, the names of
the resonances belonging to them have been suppressed.
The predicted energy at rest (equal to the  mass) 
of each level is given to its  most right.
\label{levels-nucleon}}
\end{figure}

\noindent
Almost same set of parameters, up to a modification of $d$ to  
$d= 3\, \mbox{ fm}$, fits the $\Delta (1232)$ spectrum, 
which exhibits exactly same  degeneracy patterns, and
 from which only the three $P_{31}, P_{33}$, and $D_{33}$ states 
from the $K=5$ level are ``missing''. Remarkably, none of the reported states, 
with exception of the $\Delta (1600)$ resonance, 
presumably a hybrid, drops from the systematics.
The unnatural parity of the $K=3,5$ levels requires a pseudoscalar
diquark. For that one has to account for
an $1^-$ internal excitation of the diquark which, when coupled to
its maximal spin $1^+$, can produce a pseudoscalar 
in one of the possibilities. The change of parity from natural to
unnatural can be given the interpretation of a  chiral phase
transition in baryon spectra.   
Levels with $K=2,4$  have been  attributed to 
entirely  ``missing'' resonances in both the $N$ and $\Delta $ spectra.
To them, natural parities have been assigned on the basis of a
detailed analysis of the ($1p$-$1h$) Hilbert space of three quarks and
its decomposition in the $\vert K l m, s=\frac{1}{2} \rangle $ 
basis \cite{MK-97}.
We predict a total of 33 unobserved resonances of a dominant
quark-diquark configurations  in the  $N$ and $\Delta (1232)$
spectra with masses below $\sim 2500$ MeV, much less but any other
of the traditional models.

In our previous work ~\cite{Quiry_2}, the 
potential in eq.~(\ref{ros_morse}) has been considered 
in the spirit of SUSYQM as a 
{\it central two-parameter \/} potential in $E_3$ and 
without reference to $S_R^3$, a reason for which the 
values of the parameter accompanying the
$\csc^2$ term had to be taken as integer 
{\it ad hoc\/} and for the only sake of a better fit to
the spectra, i.e., without any deeper justification.
Instead, in the present work, 
\begin{quote}
we fully recognize
that  the higher dimensional potential
${\mathcal V}(\chi, \kappa )$ in eq.~(\ref{chi_pot}),
which acts as the prototype of  Rosen-Morse I,
is a {\it non-central one-parameter\/} potential 
in which the strength of the
$\csc^2$ term, the centrifugal barrier on $S_R^3$,  is 
uniquely fixed by the eigenvalues of underlying three-dimensional
angular momentum.
\end{quote}
\end{enumerate}

\underline{\it 3.2 The nucleon spectrum in the $SO(2,1)$ symmetry scheme. }
As already announced in the introduction, the energy spectrum in 
eq.~(\ref{enrg_cot}) can equivalently be cast in terms of the eigenvalues
 of the $SO(2,1)$ Casimir, the pseudo-angular momentum operator,
${\mathcal J}^2$ \cite{Wybourne},\cite{Marcelo}. 
At first glance, it is not obvious how to re-express 
in terms of $SO(2,1)$ quantum numbers the general two-term 
energy formula in eq.~(\ref{enrg_cot}) that
contains both the quadratic and inverse quadratic eigenvalues of 
the $SO(4)$ Casimir. 
An immediate option would be
to nullify the potential strength, i.e. setting $G=0$, 
which turns one back to the free particle on
the hypersphere. In this case only the quadratic terms survives which is
easily equivalently rewritten to
\begin{equation}
E_{m^\prime} (\widetilde{\kappa})=\widetilde{\kappa }
\frac{\hbar^2}{2\mu }\left( \left(m^\prime \right)^2 -1\right),
\quad j= l+1, \quad m^\prime =j+n,
\label{G=0}
\end{equation}
where $l$ is the ordinary angular momentum label, while $n$ is the radial
quantum number (it equals the order of the polynomial shaping the wave
function labeled by $K$ in eq.~(\ref{Rom_pol}).
It is obvious that the degeneracy patterns in the
$SO(2,1)$ spectrum designed in this manner 
are same as the $SO(4)$ ones.

Perhaps nothing expresses the $SO(2,1)/SO(4)$ symmetry correspondence
better but this extreme case in which the manifestly $SO(4)$ symmetric
centrifugal energy on the $3D$ hypersphere is cast in terms of
$SO(2,1)$ quantum numbers of pseudo-angular momentum.

Although the bare $l(l+1)\csc^2\chi $ potential is
algebraically in line with   $AdS_5/CFT$ correspondence, it 
completely misses the perturbative aspect of QCD dynamics.
The  better option for getting rid of the inverse-quadratic term in 
eq.~(\ref{enrg_cot}) is to permit  $K$ dependence of 
the potential strength and choose  $G=g(K+1)$ with $g$ being
a new free parameter.
Such a choice (up to notational differences) has been suggested in \cite{Koca}
in the context of SUSYQM.
If so, then the energy takes the form 
\begin{equation}
E_{m^\prime }(\widetilde{\kappa} )=- g^2{\frac{\hbar^2}{2\mu}}   
+ \widetilde{\kappa} \frac{\hbar ^2}{2\mu }
\left( \left( m^\prime \right)^2-1\right), 
\quad m^\prime =  1, 2, 3, ....
\label{enrg_su11}
\end{equation}
The above manipulation does not affect the degeneracy patterns as it only
provokes a shift in the spectrum by a constant.
Compared to eq.~(\ref{G=0}) the new choice allows the former inverse quadratic
term to still keep presence as a contribution to the energy
depending on a free constant parameter, $g$. 
In this manner, the $SO(2,1)$ energy 
spectrum continues being described by a two-term formula, a 
circumstance that allows for a best fit to the $SO(4)$ description.

Once having ensured that the $SO(2,1)$ and 
$SO(4)$ spectra share same degeneracy patterns, 
one is only left with the task to check consistency of the
level splittings predicted by the two schemes.
Comparison of eqs.~(\ref{enrg_cot}) and (\ref{enrg_su11}) shows that
for the high-lying levels where the inverse quadratic term becomes negligible,
both formulas can be made to coincide to high accuracy by a proper choice 
for $g$.  That very  $g$ parameter can be used once again to
fit the low lying levels to the $SO(4)$ description, 
now by  a value possibly different from the previous one.

This strategy allows to make the $SO(2,1)$ and $SO(4)$ descriptions of 
non-strange baryon spectra sufficiently close and establish 
the symmetry correspondence. In that manner
we  confirm our statement quoted in the introduction that the TEC 
potential is in line  with  both the algebraic aspects of 
$AdS_5/CFT$ and QCD dynamics and provides a bridge between them.\\

\underline{\it 3.3 The proton mean square charge radius.}
In this section we shall test the potential parameters
in eq.~(\ref{parameters}) and the wave function in 
eqs.~(\ref{ros_morse}),~(\ref{Rom_pol})
in the calculation of the proton electric form-factor, the touch stone 
of any spectroscopic model. As everywhere through the paper, the
internal nucleon structure is  approximated by a
quark-diquark configuration.
In conventional three-dimensional flat space the electric 
form factor is defined 
in the standard way  as
the  matrix element of the charge component,
$ J_0(\mathbf{r})$, of the proton electric current 
between the states of the incoming, $\mathbf{p}_i$, and 
outgoing, $\mathbf{p}_f$, electrons in the dispersion process,  
\begin{equation}
G_{\mathrm{E}}^{\mathrm{p}}( |\mathbf{q}|)=
<\mathbf{p}_f| J_0(\mathbf{r})|\mathbf{p}_i>,
\quad \mathbf{q}=\mathbf{p}_i-\mathbf{p}_f.
\label{el_ff}
\end{equation}
The mean square charge radius is then defined in terms of the slope
of the electric charge form factor at origin and reads,
\begin{equation}
\langle \mathbf{r}^2\rangle =
-6\frac{\partial
G_{\mathrm{E}}^{\mathrm{p}}( |\mathbf{q}| ) }{\partial |\mathbf{q}|^2}{\Big|}_
{|\mathbf{q}|^2=0}\,.
\label{chrg_rd}
\end{equation}
On $S_R^3$, the three-dimensional radius vector, $\mathbf{r}$, has to
be replaced by, $\mathbf{\bar r}$ with $|\mathbf{\bar r}|=R\sin\chi
=\sin\chi/\sqrt{\kappa } $.
The  evaluation of eq.~(\ref{el_ff}) as four-dimensional
Fourier transform requires  the four-dimensional plane wave,
\begin{equation}
e^{i q\cdot \bar x}= 
e^{i|\mathbf{q}||\mathbf{\bar r}|\cos \theta }=
e^{i{|\mathbf{q}|}\frac{\sin \chi}{\sqrt{{\kappa}}} 
\cos \theta },
\quad |\mathbf{\bar r}|=R\sin\chi=\frac{\sin \chi}{\sqrt{{\kappa}}}.
\label{4_PW_FF}
\end{equation}
The latter refers to a $z$ axis chosen along the
momentum vector (a choice justified in elastic scattering
\footnote{A consistent definition of the four-dimensional plane wave in
$E_4$ would require an Euclidean $q$ vector. However, for 
elastic scattering processes, of zero energy transfer, where $q_0=0$,
the $q$ vector can be chosen to lie entirely in $E_3$, 
and be identified  with the physical space-like momentum transfer.}), 
and a position vector of the confined quark having in general 
a non-zero  projection on the extra dimension axis in 
$E_4$.  

 The integration volume on $S_R^3$ is given by 
$\sin^2\chi \sin\theta \mbox{d}\chi \mbox{d}\theta \mbox{d}\varphi$.
The explicit form of the nucleon ground state wave function obtained from
eq.~(\ref{Rom_pol}) in the  $\chi$ variable reads
\begin{eqnarray}
X_{(00)}(\chi ,\widetilde{\kappa } )&=&N_{(00)}e^{-b\chi }
\sin \chi ,\nonumber\\
N_{(00)}&=&\frac{4b(b^2+1)}{1-e^{-2\pi b}} \, , \quad b=\frac{2\mu G}
{\sqrt{\widetilde{\kappa}}\hbar^2}.
\label{wafu_gst}
\end{eqnarray}
With that, the charge-density takes the form,
 $J_0(\chi ,\widetilde{\kappa} )=e_p|
\psi_{\mathrm{gst}}(\chi ,\widetilde{\kappa })|^2$,
$e_p=1$. In effect, eq.~(\ref{el_ff})
 amounts to the calculation of the following integral,
\begin{eqnarray}
G_{\mathrm{E}}^{\mathrm{p}}(|\mathbf{q}|, \widetilde{\kappa } )
&=&\sqrt{\kappa} \int_0^{\pi } \mathrm{d}\chi  
\frac{
\left( X_{(00)}(\chi,\widetilde{ \kappa} )\right)^2 
\sin (|\mathbf{q}|\frac{\sin\chi}{\sqrt{\kappa }})}
{|\mathbf{q}|\sin\chi }   ,
\label{el_ff_gst}
\end{eqnarray}
where the dependence of the form factor 
on the curvature has been indicated explicitly.
The integral is taken numerically and the resulting
charge form factor of the proton is displayed in Fig.~2 together with data.

\begin{figure}[b]
\center
\includegraphics[width=7.7cm]{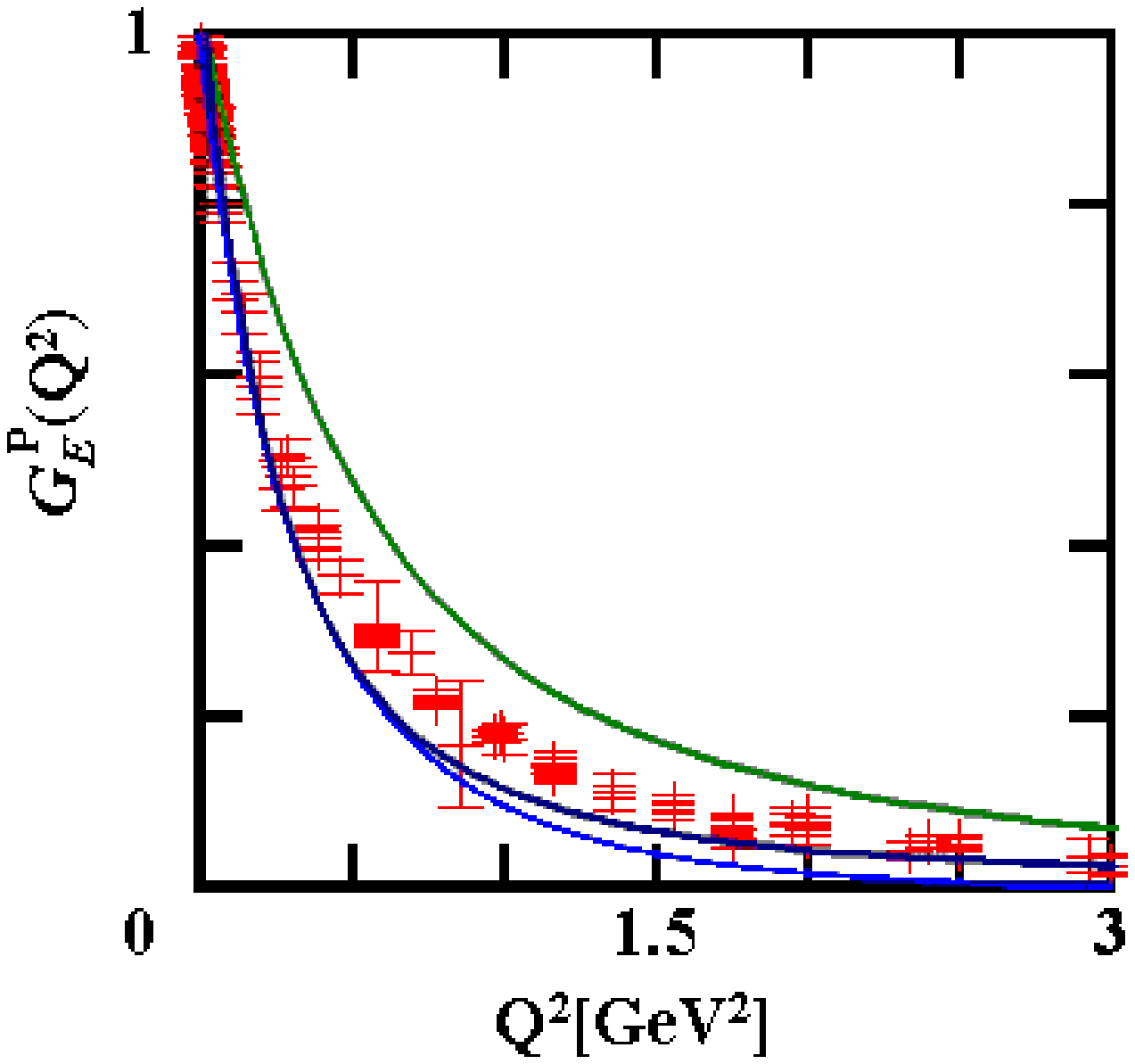}
\caption{ The electric charge form factor of the proton calculated for
various curvature parameters. The upper line corresponds to the curvature
as fitted to the nucleon spectrum, the curvature leading to the
middle line has been fitted to the experimental value of the mean square 
of the charge radius,$\sqrt{<\mathbf{r}^2>}=0.87$ fm.
The lowest line follows from a 
Bethe-Salpeter calculation based upon an instanton induced two-body 
potential and has been presented in ref.~\cite{Metsch}. 
Data compilation taken from \cite{Metsch}. 
\label{charge_FF}}
\end{figure}

\section{Curvature shut-down:The deconfinement}

The presence of the curvature parameter in the trigonometrically extended
Cornell confinement potential
opens an intriguing venue toward deconfinement as a 
$S_R^3$ curvature shut-down. It can be shown that  
\begin{quote}
high-lying  bound 
states from the trigonometrically extended Cornell
confinement potential approach scattering states
of the Coulomb-like  potential in ordinary flat space. 
Stated differently, the TEC confinement gradually fades away with vanishing
curvature and allows for deconfinement.
\end{quote}
In the direct $\kappa \to 0$ limit,
the second term of the r.h.s. in 
eq.~(\ref{enrg_cot}) vanishes and the spectrum 
becomes the one of $H$ atom-like  bound states.
In the softer $\sqrt{\kappa }K\to \mbox{k}$ (small $\kappa$, big $K$) limit,
where ``k'' is a constant, the term in question approaches the scattering
continuum. In effect, the ${\mathcal V}(\chi ,\kappa )$ 
spectrum  collapses down to the regular Coulomb-like potential, 
\begin{equation}
E_{K}(\kappa )\stackrel{\kappa \to 0}{\longrightarrow} 
-\frac{G^2}{\frac{\hbar^2}{2\mu}}   \frac{1}{n^2}
+ \frac{\hbar ^2}{2\mu }\mbox{k}^2, \quad l=0,1,2,...,n-1.
\label{bua}
\end{equation}
The rigorous proof that also the wave functions of the complete TEC potential
collapse to those of the corresponding Coulomb-like problem for 
vanishing curvature is a bit more 
involved and can be found in \cite{Vinitski}.In other words,
\begin{quote}
 as  curvature goes down as it can happen because
of its thermal dependence, 
confinement fades away, an observation that is suggestive of
a deconfinement scenario controlled by the curvature parameter
of the TEC potential.
\end{quote}
Deconfinement as gradual flattening of space has earlier been considered
by Takagi \cite{Takagi}.  Compared to \cite{Takagi}, 
our scheme brings the advantage that 
the flattening of space is paralleled  by 
a temperature driven regression of the TEC potential in eq.~(\ref{ros_morse})
to the flat Coulomb-like potential, and correspondingly,
by the  temperature driven regression of the TEC wave functions 
from the confined to the Coloumb-like wave-functions from
the deconfined phases.

\section{ Summary}

We emphasized importance of designing confinement phenomena
in terms of infinite potential barriers emerging on curved spaces.
Especially, quark confinement and QCD dynamics have been
modeled in terms of a trigonometric potential of finite range that emerges as
harmonic potential on the three-dimensional hypersphere of constant
curvature, i.e., a potential that satisfies the Laplace-Beltrami 
equation there. The potential under consideration 
interpolates between 
the $1/r$-- and infinite wells while passing through a region
of linear growth. This trigonometric confinement potential  is
exactly solvable at the level of the Schr\"odinger equation and
moreover, contains the infinite range 
Cornell potential following from 
Lattice QCD and topological field theory 
as leading terms of its Taylor decomposition.
When employed as a quark-diquark potential, it led to a remarkably 
adequate description of the $N$ and $\Delta $ spectra in explaining their
$O(4)/SO(2,1)$ degeneracy patterns, level splittings, 
number of states, and proton electric charge-form factor.
Moreover, the trigonometrically extended Cornell (TEC) potential,
in carrying simultaneously the $SO(4)$ and $SO(2,1)$ symmetries
(as the $H$ atom!),  matches the algebraic aspects of
$AdS_5/CFT$ correspondence and establishes its link to QCD potentiology.
A further advantage of the TEC potential is
the possibility to employ  its curvature parameter, 
considered as temperature dependent,  as a driver of the 
confinement-deconfinement transition in which case the wave functions of
the confined phase approach bound and scattering states of ordinary flat space
$1/r$ potential.

All in all, we view the concept of curved spaces as a promising one
especially within the context of quark-gluon dynamics.

\vspace{0.2cm}

{\Large \bf Acknowledgments}\\

\noindent
One of us (M.K) acknowledges the organizers of the
XXVII International Workshop on Nuclear Theory in the Rila
Mountains, Bulgaria, for warm hospitality and excellent working 
conditions during the event.

Work supported by CONACyT-M\'{e}xico under grant number
CB-2006-01/61286.

\end{document}